# Situation Coverage Testing for a Simulated Autonomous Car – an Initial Case Study

Heather Hawkins[1], Rob Alexander[2]

[1]AECOM, 16 Toft Green, York, YO1 6JT, UK
`heather.hawkins@aecom.com`
[2] Department of Computer Science, University of York, York, UK
`rob.alexander@york.ac.uk`

**Abstract.** It is hard to test autonomous robot (AR) software because of the range and diversity of external situations (terrain, obstacles, humans, peer robots) that AR must deal with. Common measures of testing adequacy may not address this diversity. Explicit situation coverage has been proposed as a solution, but there has been little empirical study of its effectiveness. In this paper, we describe an implementation of situation coverage for testing a simple simulated autonomous road vehicle, and evaluate its ability to find seeded faults compared to a random test generation approach. In our experiments, the performance of the two methods is similar, with situation coverage having a very slight advantage. We conclude that situation coverage probably does not have a significant benefit over random generation for the type of simple, research-grade AR software used here. It will likely be valuable when applied to more complex and mature software.

## 1   Introduction – it is Hard to Test Autonomous Robots Thoroughly

It is hard to test autonomous robot (AR) software because of the range and diversity of external situations (terrain, obstacles, humans, peer robots) that the robots must deal with. AR must interpret a wide variety of stimuli from their world and make decisions that are appropriate given the specific combination of stimuli they are receiving – their relationship with their environment is complex. They must also deal with the reactive nature of their environments; for example, other vehicles in the environment will react to the AR's actions, which will in turn change the stimuli that the AR receives, and thus its own subsequent behaviour [1]. Many features of the environment that matter are not specific to the mission goals – they are simply present, and must be handled [2]. Beyond merely needing to avoid accidents, AR may need to comply with detailed-yet-ambiguous rules intended to guide human behaviour (such as the UK Highway Code and Rules of the Air).

If AR are tested using a *system coverage* approach – if they are tested until some set of system elements (e.g. software functions, code paths, code branches) have all been tested at least once – this will not necessarily identify all faults. Many potential faults are of the type "fails to deal with situation X", and it may be that no situation component has been designed to "deal with situation X". For example, if an AR's motion control components have no logic for handling "overtaking an over-wide vehicle" then system coverage will not guide testing to identify faulty behaviour in that situation.

If AR are tested using a *requirements coverage* approach – if they are tested until their conformance with all identified requirements has been adequately assessed (for some definition of "adequately") – this will not necessarily identify all issues. Some external situations may not have been identified as requirements for handling – this represents a failure of validation (with respect to reality) rather than verification (of system against specification). Requirements-based testing is primarily a verification activity, so might well not find such faults. For example, if there is no explicit requirement to handle "overtaking an over-wide vehicle", then requirements coverage will not guide testing to identify faulty behaviour in that situation.

If AR are tested using a s*cenario coverage* approach – where the aim is to cover a representative set of scenarios described by linear sequences of stimuli to the AR – this will not necessarily identify all issues. It is likely that a relatively small set of scenarios will be explored, and that these will be explored in a narrow way. Adequate testing of AR requires that we study not just linear scenarios but the dynamics that arise from the interaction of the AR with their environment. Much of the interesting behaviour here happens once the AR are off the nominal scenario course and having to respond to challenging circumstances.

The authors have previously identified *situation coverage* [1] as an important technique for testing AR. Put simply, situation coverage is a measure of the proportion of all possible situations (that the software under test could conceivably encounter) that have been tested by some test set. Unlike linear scenarios, situations here are starting states and rules for projecting future states – they do not commit to a linear sequence of events. Like any coverage criterion, situation coverage can be used to assess the adequacy of a test set, and to guide automated test generation.

Situation coverage is a promising criterion, but there is very little empirical work that evaluates it. In this paper we empirically explore a key research question – *"Can a situation generation testing method guided by a situation coverage measure outperform a randomly-driven one, in terms of diverse faults detected, at the same level of computing effort?"* Specifically, we assess this for a multiple-subsystem simulated AR.

## 2 Situation Coverage is a Promising but Under-Evaluated Technique for AR Validation Testing

### 2.1 Situation coverage has potential

Situation coverage has the potential to be both practical and effective. At the simplest level, if we have a concept of situation coverage, and metrics that measure it, then we have grounds for judging whether we are exploring a sufficient diversity of situations. As we expand our diversity of situations, there is the potential to reveal an increasing number and type of faults. Some of these faults will be simply in implementation of the AR's specification, but others may be in the specification itself. Testing driven by situation coverage may thus have not just a verification role but a validation one as well.

To take a simple example, imagine a wheeled AR which was designed without explicitly considering the multitude of interactions that could take place at a box junction. It may have a very detailed driving behaviour specification, but it is deficient in this respect. Its behaviour at such junctions may be adequate in most cases (because of its general-case driving rules), but there may be rare corner cases in which it is dangerous. If a situation coverage measure is used that requires a wide range of junction situations to be run, it may reveal the gaps in the specification.

Indeed, merely enumerating the components of situations that can be encountered (terrain, weather, peer actors, "missions" …) may itself cause analysts to consider more cases. Consider the typical experience when formalising requirements – in the process of making the requirements precise, inconsistencies and ambiguities in them are brought to light.

A common concern when defining tests is that the requirements errors made by the testers (e.g. invalid assumptions, misunderstandings of intent, superficial models of external phenomena) may be the same set of requirements errors made by the programmers – the tests may thus have blind spots exactly where the programs have faults (cf. the Knight-Leveson experiment [3]). Defining a situation space, and then covering it by generating situations, is a very indirect way of defining tests, so there is a relatively high chance that analysts will *not* make the same wrong assumptions that developers made when designing the system. This is not to say that the analysts will make fewer mistakes, merely that they are less likely to make the same ones.

There are already existing taxonomies on which situation coverage measures can be built. Dogramadzi et al, in their paper on Environmental Survey Hazard Analysis [2], provide a partial taxonomy of environmental factors that need to be considered for AR safety. Similarly, Woodman et al [4] provide a structured list of entity types, properties, and actions for home and urban environments.

Once a situation space is defined, individual situations need to be generated. In theory we could hand-craft these directly into a simulation model, but the effort this involves is likely to be impractical. We will need to automatically generate situations, each one representing a position in the situation space. Given a good metric, automated testing can work very well. Without a metric, such techniques are often quite limited – no metric means no heuristic search or similar techniques, so we are reduced to random (or brute-force exhaustive) test generation.

The level of rigour we suggest here is perhaps best related to that of DO-178C [5] and similar – we are proposing rigorous testing using a coverage measure (analogous to 178's use of MCDC coverage for software), not the kind of absolute completeness that formal methods often strive for. Such rigour makes situation-coverage-based testing potentially useful evidence in a safety case.

We are not suggesting that anyone should use *only* situation coverage to guide their testing. Relying on a single form of coverage criteria is unwise, and relying wholly on coverage criteria for test generation is very unwise (on the latter, see Gay et al. [6]).

### 2.2 Situation coverage is inadequately empirically explored

There is a little existing work on situation coverage, but only in narrow domains. For example, the VITRO approach [7] is only concerned with static scenes and non-interactive videos for testing computer vision.

In particular, there is not much empirical evaluation of testing driven by situation coverage. Andrews et al [8–10] explore several coverage criteria over their situation spaces – some of their criteria make tractable test sets, but they perform no evaluation of actual fault-finding power.

Similarly, Zendel et al [7] use low geometric discrepancy sampling to make tractable subsets of situations, but they don't rigorously evaluate their testing power. It is thus not possible to assess how much confidence we can have in the results of such testing. Zendel et al [11] generate test cases and apply them to test software, but do not systematically explore their ability to detect diverse faults (nor do they apply the test cases to diverse SUTs – they only use one, and that is a vision subsystem, not rather than a whole AR).

## 3 We Have Implemented a Simple Example of Situation-Coverage-Driven Testing

To pursue the research question given Section 1, we implemented a simple simulation of an autonomous car, along with a test engine that generated test situations either (i) randomly or (ii) guided by situation coverage.

### 3.1 Our simulated world represents some real challenges

The simulated world is simple, but it is a representative simulation of a plausible road network with typical obstacles encountered therein. An instance of the simulated environment – which constitutes "one situation" – consists of a bounded world map containing the following static objects: straight-line roads, junctions (T or dead-end), parked car obstacles and a *Target* location. The map also contains a number of dynamic objects, including moving car obstacles and the AR itself. Moving obstacles can choose, at random, to change direction at a junction. The number of moving obstacles on the road network is kept roughly static – they can leave the network at dead-ends, but then are replaced by a new moving obstacle in a dead-end chosen at random.

The task required in each situation is for the AR to navigate around the road network to reach its target, whilst not crashing into obstacles, leaving the road surface, or moving inappropriately into the opposing lane. Unlike the moving obstacle cars, the AR is not permitted to leave the network at dead-ends – it must U-turn and continue searching.

Within the simulation space provided by the features listed above, we can automatically generate a wide variety of such situations. Maps are generated programmatically, randomly adding junctions and road connections between them, using a random number generator to choose how many junctions to add to the map. Various rules constrain the map configuration e.g. there is a minimum junction separation, and roads must not extend beyond map boundaries. Random choice also governs the number of obstacles (moving and static) and their locations, and the start location of the AR and the location of its Target. Once again, rules constrain the choice to prevent e.g. the AR being added at the same location as an obstacle, or parked cars overlapping. Each time it generates an element, the program has a limited number of chances to randomly place each element; if a suitable location cannot be found given the permitted effort then it is omitted from the map.

The generator uses a seed, here called the "external seed". If the external seed is known, the starting configuration of the map – including AR and Target locations – can be regenerated. A separate "internal seed" seeds the random number generator used for random decisions during the same run. These decisions include where the AR first turns at a new junction, and where moving car obstacles re-join the road network after leaving. If both seeds are known, a given run can be regenerated exactly.

The simulation is able to detect a range of unwanted "accident" events occurring during the simulation, including:

- CLASHWITHOBSTACLE – collision between the AR and a parked car
- CLASHWITHOTHERCAR – collision between the AR and a moving car
- LEAVEROAD – AR leaves the driveable part of the road
- CROSSCENTRELINE – AR crosses the into the other lane without justification

We implemented the simulation in MASON, a Java library for multiagent simulation [12]. We chose this, rather than a more complete and sophisticated robot or vehicle simulation tool, so that we would have control over exactly how the simulation worked.

The screenshot in **Fig. 1** shows the visualisation of the simulated world for one particular map.

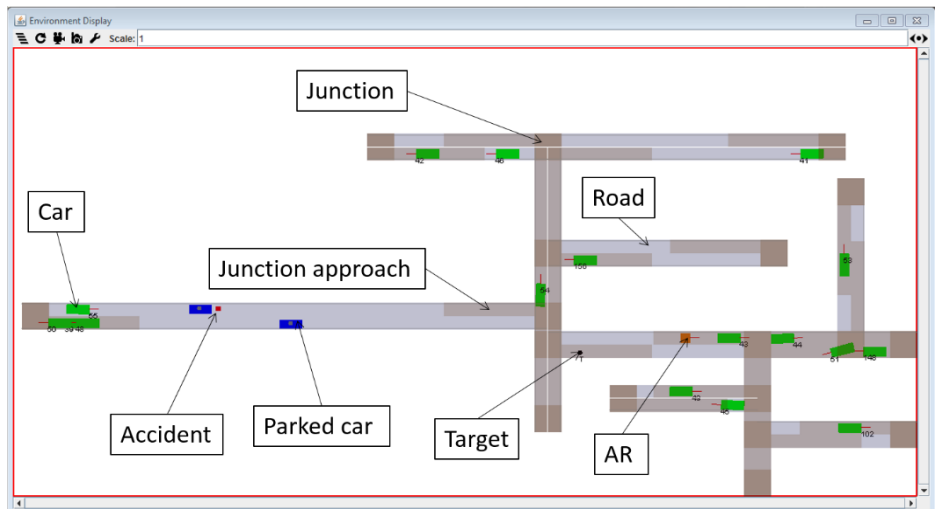

**Fig. 1.** - Simulation visualisation

### 3.2 Our simulated autonomous car is a plausible target for testing

Our simulated vehicle is simple, as this is sufficient for its simple environment. But it contains features that are representative of some of the complications that a comparable real vehicle would experience. For example, it simulates visual following of road markings and LIDAR detection of obstacles. It is unable to sense through obstacles, so attempts to predict whether a second parked car immediately follows the first. It also attempts to predict the trajectories of oncoming vehicles.

Navigation is stochastic, although the AR favours roads that it has not visited previously, with a preference for those which will take it closer to its Target (in practice this would require the AR being pre-programmed with the relative location of its Target). Moving obstacles can choose, at random, to effect a direction change at a junction.

The AR controller software contains a number of basic rules to facilitate safe behaviour, although the complexity inherent in even this simple simulation environment means that there are occasions when the rule base is insufficient to accommodate the precise situation which the AR is presented with, and thus failures may occur.

The AR controller is seeded with a range of faults, 20 in all. Each is a plausible software problems for a system like the SUT, and all were relatively easy to implement in a small-scale project. They do not necessarily provide good coverage of the space of possible or likely faults.

Each fault can be toggled on or off for a given run, and if a fault is triggered during a run (i.e. its code is actually executed) this is logged. Faults include reduced sensor range, partial sensor coverage, and systematic errors in transforming sensor data to internal map changes.

### 3.3 Our situation coverage criterion has potential to lead to diverse situations

We have situation coverage criteria, specific to the environment and AR task modelled by our simulation, which are likely to lead to a diverse range of situations being generated. The three criteria we used are:

- *Distance Previous Junction to Target* – this ensures that the AR encounters the target in both immediately-post-junction and normal driving modes
- *Minimum Distance Target to Obstacle* – this ensures that the AR encounters the target in both near-obstacles and open road modes
- *Distance AR start to Target* – this ensures that the AR experiences runs where the target is immediately accessible and runs where the target requires a long journey to reach

We chose the criteria mostly because they were easy to conceptualise and implement. They are not ideal. Each presents a continuum of one parameter, rather than a space of qualitatively different possibilities (e.g. different driving modes for the non-AR cars). They are all distances, rather than any of the other property types that could be used. They are all quite indirect, in that any given value can be satisfied by many, many different maps. They nevertheless have the potential to lead to many different situations, particularly at the extremes of their values, and thus are suitable for this initial exploration.

To combine the criteria, we discretize each to an (arbitrary) six levels, where the extremes of levels 0 and 5 are the minimum and maximum, respectively, of the corresponding measure. This discretization gives us a situation space that can be filled progressively by generated situations. Specifically, we treat the three discretized criteria as the axes of a 3-dimensional space of (6x6x6=216) possible situations.

The discrete locations within this situation space are effectively equivalence classes. Each implies a uniformity hypothesis – that all situations in that class will cause the same fault behaviour of the SUT.

The combination of these three criteria gives us a reasonable range of required situation properties. They all focus on the context of the target, as the area around the target is the most complicated part of the maps for the AR to navigate, having all the features of other road segments and the complication of having to steer to reach the target itself. In a real application, there would be far more than 216 possible situations; the limited set we use here kept our experimental work tractable in a small project.

### 3.4   Our test generation approach is driven by the coverage criteria

We implemented a simple search approach that efficiently achieves coverage-driven testing.

In theory, given a well-defined finite situation space, particularly a small one as we are using here, one can exhaustively generate runs by walking through the locations in the space and generating situations (here, maps) accordingly. Building a generator that takes such a situation outline and produces a corresponding map deterministically is, however, very difficult.

Instead, our test driver generates test situations randomly, as described in Section 3.1. To exploit the situation coverage criteria for guidance, it uses a simple technique common in coverage-driven verification of hardware [13] – after generating each map, it checks it against the situation space model described in Section 3.3. If the corresponding cell (from the set of 216) is unfilled, it runs the map in simulation; if the cell is already filled, it discards the map and generates another one. It terminates when a preset number of maps have been generated and checked.

### 3.5   Our code is available

The full code for the simulation, vehicle and test generator can be found at http://dx.doi.org/10.15124/1b0286c0-ceaf-4c70-8ec1-7895f090079e, along with basic documentation.

## 4 Experimental Evaluation

We carried out an experiment to evaluate how successfully the test generator described in section 3.4 discovers seeded faults in the simulated AR described in section 3.2.

### 4.1 Experimental setup

As a standard of comparison, we implemented a simple random test generator – essentially the coverage-driven generator without the coverage-filter step. This generator, as a consequence, will run every map that it generates. In the extreme case, all tests it runs might fit into the same coverage criteria cell.

The complexity of our SUT's behaviour, and the relative immaturity of its codebase, meant that there was a relatively high rate of background (non-seeded) faults. Of the 20 faults seeded into the SUT, we selected 7 that were relatively easy to trigger, and thus would be reliably detectable above the noise of the background faults. We will discuss in Section 5.2 some consequences of this.

The set of faults we used are shown in **Table 1**. They correspond to a range of plausible hardware or software problems for an AR.

Table 1. Faults used in the experiments

| ID # | Description |
|---|---|
| 2 | When creating a waypoint (a point which the AR will steer towards), misplace it slightly northeast |
| 4 | When creating a waypoint, misplace it slightly southeast |
| 8 | Remove the first half of the scan arc for road markings |
| 10 | Halve visual resolution when searching for road markings |
| 12 | Halve the range of the scan for road markings |
| 17 | Fail to keep turning towards the current waypoint when overtaking |
| 18 | While overtaking, look for obstacles only in direction of current heading, not in intended direction of travel |

### 4.2 Experiments conducted

We ran an experiment to test the research question from Section 1 – *"Can a situation generation testing method guided by a situation coverage measure outperform a randomly-driven one, in terms of diverse faults detected, at the same level of computing effort?"* Specifically, we ran our coverage-guided testing method for 20000 candidate maps, and simulated any map that would fill a new coverage cell. Maps that did not fill a new coverage cell were discarded. When we simulated a map, we ran it seven times – once with each of the seven faults listed above. For statistical confidence, we repeated this whole process five times. On average, each replication achieved around 80% situation coverage (i.e. around 170 of the 216 coverage cells filled).

On each repetition of the above, we recorded the length of time taken for the above generation, filtering, and simulation process, and allocated the same amount of compute time to our random comparison generator. This generator produces a map at random and then runs it against each of the seven faults in turn, before generating another map at random; there is no pre-filtering process, so all maps generated are run. On average, across the five repetitions, the random generator simulated 218 maps and achieved 36% situation coverage (i.e. around 78 of the 216 cells filled).

Generally, the random method ran more maps in the time than the coverage-guided method did. This appears to be because the coverage-guided maps had a longer average distance from AR start to Target location. This was partly driven by the "*Distance AR start to Target*" which required large distances to cover the high end of its range. Purely random maps did not have this selection pressure, so tended to be smaller.

We assessed the performance of the two approaches using three measures:

- The measure "method-prop-fault" was the proportion of the seven seeded faults that each method revealed across all the maps that it ran. This is a clear measure of overall testing power – the proportion of the faults we seeded that can be detected by the testing approach at all.
- The measure "prop-map-all-fault" measured the proportion of maps that revealed all seven seeded faults. It is a measure of map quality, i.e. of the ability of generated maps to detect many faults. It is of course only applicable if the previous measure is 1 (one – all faults were detected between all maps run by the method), and then may still be zero (if no single map revealed all seven faults). Given our argument for situation coverage given in sections 1 and 2, we would expect that in a realistic testing situation this would be zero i.e. no single map can reveal all faults. After all, if a single map (situation) can reveal all faults, then there is no need for a diversity of maps; we would do better to search for the single best map.
- The measure "avg-map-fault" measured the number of faults revealed by each map on average. It is another measure of map quality, slightly more sensitive than the previous one.

Our null hypotheses (one per measure) are that the two methods give the same results. Our corresponding alternate hypotheses are that the coverage-driven method gives a higher score.

### 4.3 Results – situation-coverage-driven testing was not significantly better

The results of our experiments for method-prop-fault and prop-map-all-fault are summarised in **Fig. 2**. The corresponding raw data can be downloaded at http://dx.doi.org/10.15124/1c787c9f-2860-46db-a323-242d8fe9aeb1.

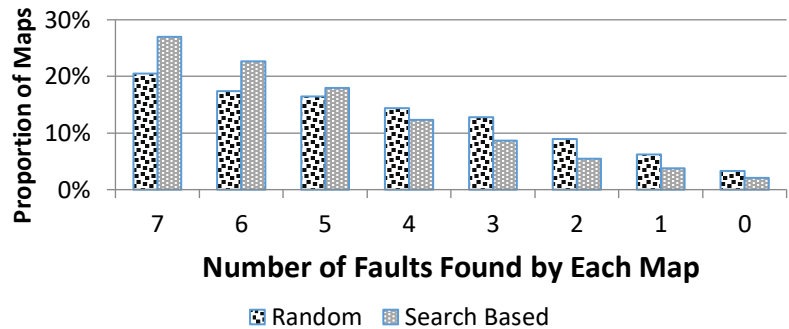

**Fig. 2.** - Seeded faults found by each map

We can see from the figure that for both methods some maps found all seven faults, so for both methods method-prop-fault is 1 (one). This is perhaps not surprising, This shows that the two methods do not differ in their ability to find faults of the types that we have seeded, given a fairly large compute budget for the low complexity of the SUT and the simulated world.

We can also see that the methods differ slightly in their ability to find all seven faults in a given map - prop-map-all-fault is thus ~0.20 for random and ~0.27 for coverage-driven. Thus the average map quality is slightly higher for coverage-driven testing.

The avg-map-fault is 4.50 for random and 5.05 for coverage-driven. This, again, suggests slightly higher map quality for a coverage-driven approach.

If we look at how the two approaches differ in terms of detecting specific faults, as shown in **Table 2**, we can see that faults 17 and 18 are relatively difficult for random testing to detect. This is probably because the faults are only triggered during overtaking, which is more likely to occur if the AR has to travel a long way around the network to search for the target. In a simple map, the AR may not have to overtake at all.

**Table 2.** - Relative proportion of runs finding each fault

|         | Difference between % of runs with faults found in SB and R runs ||||||||
|---------|---|---|---|---|---|---|---|
| Fault # | 2 | 4 | 8 | 10 | 12 | 17 | 18 |
| Averages | 5% | 4% | 5% | 3% | 6% | 16% | 16% |

Although the results do not show a (practically) significant difference in performance, the search-based map sets did appear to be of higher quality than the random map sets. This warrants further exploration.

## 5 Why Was Performance Worse than We Expected?

The results above are not what we hoped for. We have investigated why.

### 5.1 Coverage-driven testing did indeed achieve coverage

**Fig. 3** shows how situation coverage varies with number of candidate maps generated (or the equivalent compute time) for the two methods. It is clear coverage-directed generation does cover a greater proportion of the situation space in the same amount of compute time. This is what we would expect, and is compatible with the superior performance we had hoped coverage-driven testing would provide.

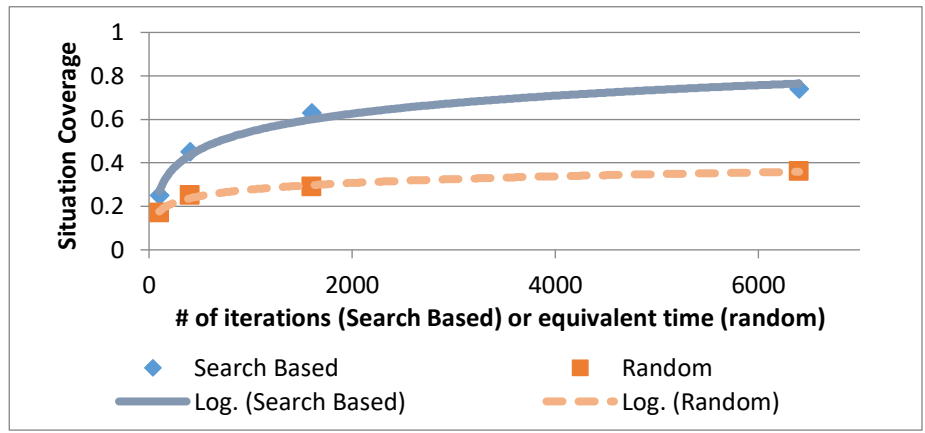

**Fig. 3.** – Situation coverage by search effort

### 5.2 There was only weak correlation between coverage achieved and number of faults found

We established in Section 5.1 that coverage-driven testing does achieve high situation coverage. This does not, however, seem to translate to discovery of faults. We ran a series of experiments that used coverage-driven testing at a range of effort levels (between 200 and 2500 candidate maps) to achieve a range of situation coverage levels (between 38% and 68%). 5 map sets were generated for each effort level, and each map set was run with each of the seven faults seeded in turn. **Fig. 4** shows the percentage of runs in each experiment in which a fault was found (was triggered and apparently led to an accident).

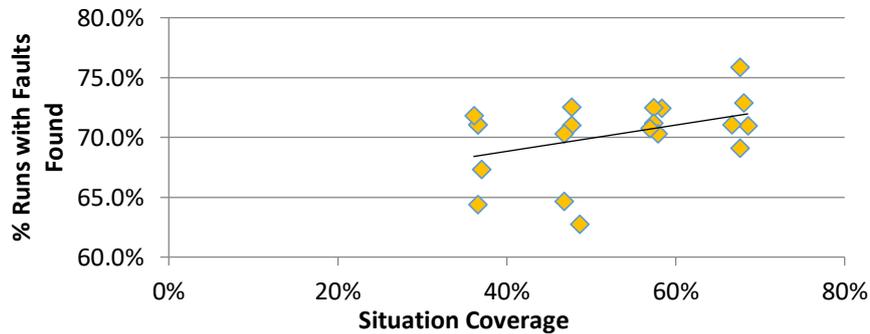

**Fig. 4.** – Faults found at different levels of situation coverage

Note – two map sets in the "38% effort" category produced exactly the same results (37% coverage, 71.1% of runs finding at least one fault), so they appear as a single point on the diagram.

One can see from the figure that correlation between situation coverage and % of runs that found faults is poor (the linear regression shown in **Fig. 4** has $r^2$=0.1681). In other words, it is not clear that improving situation coverage actually improves the number of faults found.

### 5.3 The set of faults we used may have obscured the difference between methods

As noted in Section 4.1, we used a subset of our seeded faults that were easy to detect. Only these strong-signal faults were detectable above the noise from non-seeded "endogenous" faults – even when no faults were seeded, around a third of runs featured at least one accident event. The endogenous faults gave a high probability of a false positive, where a seeded fault was triggered, this did *not* lead to an accident, but the system logged an accident for that fault because there was an endogenous-fault accident in the same run. By the measures from section 4.2 the fault would appear found, but had it been a real fault it would not likely have been found by any study of that run.

If it is true that the faults were indeed easy to find, then it is no surprise that coverage-driven testing did little better than random testing, as random testing was already doing very well indeed. In the main experiment reported in Section 4, random testing was sufficient to find all faults, with 20% of maps being able to reveal all faults.

A solution to this would be to seed harder-to-find faults. In particular, faults that rely on several properties of a situation or require a specific sequence of events are notoriously hard to identify by random testing. To do this, however, we would need to use a more mature SUT. This would exhibit fewer endogenous faults, meaning that almost all occurring failures would be caused by whatever faults were currently seeded.

Given that higher level of maturity, a more complex SUT and environment would also likely make it easier to seed (or find and capture) more complex faults. We can therefore conjecture that more selective situation generation techniques will be more effective on such systems.

## 6 Conclusions

We argued in Section 2 that the idea of driving autonomous robot testing by situation-coverage-driven situation generation is a promising one, and has potential to complement other forms of testing. Our coverage-driven implementation as described here does *not* achieve significant benefits over random situation generation, at least in terms of the SUT and fault set we have used to evaluate it. The small benefit it does show suggests further investigation may be worthwhile.

Through analysis of our experiments and results we have identified how improvements could be made that would likely improve coverage-driven performance and thus increase the gap between that and random testing. We have made source code for the test tool, simulation and AR model available for any other researchers who wish to attempt this.

From the results and our analysis, we can conclude that testing driven by situation coverage is most likely to be effective for mature, complex SUTs paired with reasonably sophisticated environment (and hence situation) models. Future work applying it in such contexts would be very valuable.

## Acknowledgements

We would like to thank Susan Stepney, John Clark, Tim Kelly, Alan Winfield, John McDermid, Kirsten Eder, Richard Hawkins and Ibrahim Habli for their contributions to our understanding of this topic, and Robert Lee and Xueyi Zou for their work on earlier versions of the simulation. Alexander and Hawkins were funded to work in this area by EPSRC grant EP/L00643X/1 (the latter while employed at the University of York).